\documentclass[11pt,a4paper]{article}

\usepackage{amsmath, amssymb, graphics, amsthm}
\usepackage{epsfig}
\usepackage{color, xcolor}
\usepackage{fancyhdr} 
\usepackage{manfnt}
\usepackage[utf8]{inputenc}
\usepackage[T1]{fontenc}
\usepackage{enumitem}

\usepackage{hyperref}

\usepackage[normalem]{ulem}

\makeatletter
\@addtoreset{equation}{section}
\makeatother

\newcommand{\be}{\begin{equation}}
\newcommand{\ee}{\end{equation}}
\newcommand{\bse}{\begin{subequations}}
\newcommand{\ese}{\end{subequations}}
\newcommand{\ba}{\begin{eqnarray}}
\newcommand{\ea}{\end{eqnarray}}

\title{{\sf Observer's observables. Residual diffeomorphisms}}
\author{
{\sf P. Duch}$^{1}$\thanks{{\sf 
pawel.duch@uj.edu.pl}},
{\sf J. Lewandowski}$^{2}$\thanks{{\sf 
jerzy.lewandowski@fuw.edu.pl}},
{\sf J. \'Swie\.zewski}$^2$\thanks{{\sf 
swiezew@fuw.edu.pl}}\\
\\
{\sf$^1$ Institute of Physics, Jagiellonian University, \L ojasiewicza 11, 30-348 Krak\'ow, Poland}\\
{\sf$^2$  Faculty of Physics, University of Warsaw, Pasteura 5, 02-093, Warsaw, Poland}
}
\date{{\small\sf \today}}

\begin{document} 

\setcounter{tocdepth}{2} %-- part,chapters,sections, subsections

\maketitle

{\sf
\abstract{We investigate the fate of diffeomorphisms when the radial gauge is imposed in canonical general relativity. As shown elsewhere, the radial gauge is closely related to the observer's observables. These observables are invariant under a large subgroup of diffeomorphisms which results in their usefulness for canonical general relativity. There are, however, some diffeomorphisms, called residual diffeomorphisms, which might be `observed' by the observer as they do not preserve her observables. The present paper is devoted to the analysis of these diffeomorphisms in the case of the spatial and spacetime radial gauges. Although the residual diffeomorphisms do not form a subgroup of all diffeomorphisms, we show that their induced action in the phase space does form a group. We find the generators of the induced transformations and compute the structure functions of the algebras they form. The obtained algebras are deformations of the algebra of the Euclidean group and the algebra of the Poincar\'e group in the spatial and spacetime case, respectively. In both cases the deformation depends only on the Riemann curvature tensor and in particular vanishes when the space or spacetime is flat.
}
}

\tableofcontents

%~~~~~~~~~~~~~~~~~~~~~~~~
\section{Introduction}
%~~~~~~~~~~~~~~~~~~~~~~~~

A crucial challenge of the canonical formulation of any generally relativistic theory remains to propose Dirac observables, i.e.\ gauge invariant functions, which would be useful in applications. The prime example of such a theory is general relativity with its gauge freedom being parametrized by diffeomorphisms. Although the search for diffeomorphism invariant observables lasts for almost a century, there still seems to be confusion in the subject and claims ranging from denying existence of any observables (except usually for the ADM charges which are defined only in asymptotically flat cases) to claiming that the problem has been solved entirely appear in the literature.

Certainly, if one allows some restrictions on applicability or some additional input then there is a number of examples of families of observables one can construct, e.g., see \cite{Komar:PNAS, PhysRevLett:4:432, RevModPhys:33:510, GeheniauDebever} for observables which work as long as no local symmetries are present, \cite{Kijowski1990, KucharTorre1991, RovelliSmolin1994, BrownKuchar1995, BicakKuchar1997, Giesel:2012rb} for examples involving additional matter content with varying level of physical relevance. Moreover, there are important general results, see \cite{Torre:1993fq} for arguments against existence of locally defined observables or \cite{Dittrich:2004cb, Dapor:2013hca, Dittrich:2005kc} for an abstract, general scheme of defining observables based on the relational approach.

A problem related to the diffeomorphism invariance is that of an evolution of gauge invariant observables.  Here, by now, the situation seems to be clear with a possible solution offered by the relational approach \cite{Rovelli:1990jm, Rovelli:1990pi, Rovelli:1989jn, Rovelli:1990ph}, being realized in time-deparametrization schemes (e.g. \cite{KucharTorre1991}).

Since it is impossible to find observables with all the desired properties (namely, for them to be defined with no additional content of the theory, regardless of any symmetries, to be locally defined, that is depending on canonical data up to finite derivatives at a point, and be suitable for canonical treatment) we are forced to look for compromises.

A recently proposed family of observables are the observer's observables \cite{Duch:2014hfa}. They require an introduction of an object referred to as the observer (possessing a location and a way of distinguishing directions), but offer a treatment of the spatial diffeomorphism freedom. The time reparametrizations are dealt with separately, e.g., with the use of an irrotational dust \cite{KucharTorre1991}. In \cite{Duch:2014hfa} it has been shown that an important virtue of this proposal is that their variations and hence Poisson brackets are not only under explicit control, but their Poisson algebra is particularly simple. Namely, there is a canonical, local subalgebra of observables which can be used to parametrize the whole gauge invariant content of the theory \cite{Duch:2015qxa}. This has been translated to the language of gauge fixing and investigated as the imposition of the radial gauge in \cite{Bodendorfer:2015:RG1}. It has been noticed there that the construction is particularly suitable (and simplifies vastly) for spherically symmetric setting. This fact allowed a novel treatment of spherical-symmetry reduction in the context of loop quantum gravity \cite{Bodendorfer:2014wea, Bodendorfer:2016:RG2}.

In the current article we investigate the family of spatial diffeomorphisms not fixed by the radial gauge. Since the introduction of that gauge is based on the presence of a non-dynamical observer, there are so-called residual diffeomorphisms which preserve this gauge, and hence need to be addressed separately. In the language of observables those diffeomorphisms correspond to transformations between different observers, so they may be of importance for some applications of the formalism.

We would like to also note here, that a change in the subtleties of the definition of the observer's observables leads to a construction in which not only spatial but spacetime diffeomorphisms are addressed. The gauge-fixing counterpart of this construction being the spacetime radial gauge\footnote{Also known in the literature as Fefferman-Graham, axial or holographic gauge.} has become of interest in the context of the AdS/CFT conjecture \cite{Heemskerk:2012np, Kabat:2013wga, Donnelly:2015hta} as a possible scheme of dealing with the diffeomorphism freedom in the bulk. However as shown in \cite{Bodendorfer:2015kua}, the algebraic structure of the spacetime radial gauge (namely the corresponding Dirac brackets) possesses some nontrivial features limiting its potential applicability severely.

The article is organized as follows. We start the section \ref{spatialcase} by recalling the crucial steps of the construction of the observer's observables which sets the stage for the current results. Next, we identify the residual diffeomorphisms and argue that they do not form a group due to the lack of closure of their action. However, when investigating the action those diffeomorphisms induce in the phase space, we notice that their generators do form a Lie algebra. Moreover, we find explicitly the structure constants (or in this case functions) of that algebra. In section \ref{spacetimecase} we present an analogous result in the case of the spacetime variant of the observer's observables construction.  Appendix \ref{app:prop_norm_coor} contains some identities fulfilled by the metric in the normal coordinates (which we call adapted coordinates) whereas in Appendix \ref{appendixdrugapochodnaZ} we include derivations of some technical results mentioned in the main text. Appendix \ref{ZinDiracmatrix} is devoted to drawing a connection between the vector fields studied in section \ref{spacetimecase} and previous work on the subject and proves a uniqueness property mentioned in the main text.

We use the following conventions regarding the order of the indices in the Christoffel symbol and the Riemann tensor
\be
	\nabla_XY^a = X^b\partial_bY^a + \Gamma^a{}_{bc}Y^bX^c, \qquad [\nabla_X,\nabla_Y]Z^a = R^{a}{}_{bcd}Z^bX^cY^d
\ee
assuming $[X,Y]=0$.

%~~~~~~~~~~~~~~~~~~~~~~~~
\section{Residual diffeomorphisms in the spatial case}\label{spatialcase}
%~~~~~~~~~~~~~~~~~~~~~~~~

We start by recalling the steps of the definition of observer's observables as they were defined in \cite{Duch:2014hfa} crucial for the current purposes. We will work in a canonical formulation of a generally relativistic theory, i.e., in a phase space $\Gamma$ with points labeled by the spatial metric tensor $q$ of a 3-dimensional spatial slice $\Sigma$
%\footnote{We assume it has trivial topology to simplify global considerations.} 
of the spacetime $\cal M$, its canonically conjugate momentum $p$ being a tensor density and possibly also some matter fields. The matter fields will not be of relevance for the presented results so we focus on the gravitational sector of the theory.

%~~~~~~~~~~~~~~~~~~~
\subsection{Observer}\label{sec:defObserver3d}

In this theory we introduce an observer ${\cal O}$. It is specified by
\begin{enumerate}[label=\it{(\roman*)}]
\item a point $\sigma_0\in\Sigma$, representing the location of the observer and 
\item a map assigning an orthonormal tangent frame at $\sigma_0$ to each metric tensor $q$ on $\Sigma$\footnote{In \cite{Duch:2014hfa} this second constituent of an observer has been described as a fixed tangent frame $e^0_I\in T_{\sigma_0}\Sigma$, from which a metric dependent frame is produced using a Gram-Schmidt orthonormalization process with respect to the given metric $q$. For our current purposes, however, we can leave the details of the choice of the orthonormal frame unspecified, as the results we obtain in what follows do not depend on that choice (for a discussion of this issue see section \ref{sectionComments}).}
\be
	q\ \mapsto\ e_I\in T_{\sigma_0}\Sigma,
\ee
where $I\in\{1,2,3\}$ is a label.
\end{enumerate}
A single observer
\be
	{\cal O} = (\sigma_0,e_I(\cdot))
\ee
is fixed throughout this section. Auxiliary observers\footnote{Note that, to comply with the notation from \cite{Bodendorfer:2015:RG1}, we denote ``a different copy of'' with a bar rather then the more standard apostrophe.} $\bar{\cal O}, \bar{\bar{\cal O}},\ldots$ will be considered to construct examples of residual diffeomorphisms later on.

%~~~~~~~~~~~~~~~~~~
\subsection{Observer's internal space of labels}

The observer ${\cal O} = (\sigma_0,e_I(\cdot))$ probes $\Sigma$ and the data thereon using a metric dependent map 
\be\label{exp}
	\xi:\mathbb{R}^3\ni (x^I)\ \mapsto\ \exp_{\sigma_0}(x^Ie_I)\in\Sigma,
\ee
where $\exp$ denotes the exponential map, which means it assigns to each triple of numbers $(x^1,x^2,x^3)$ the point $\gamma_{\sigma_0,x^Ie_I}(1)$, where $\tau \rightarrow \gamma_{\sigma_0,x^Ie_I}(\tau)$ is a geodesic segment with respect to metric $q$ defined in $\Sigma$ by its starting point $\sigma_0$ and the initial velocity vector $x^Ie_I\in T_{\sigma_0}\Sigma$. This assignment of $(x^I)$ to points in $\Sigma$ is what was referred to as Cartesian adapted coordinates in \cite{Duch:2014hfa}. We will also refer here to the spherical adapted coordinates $(r,\vartheta,\varphi)$ being related to the above ones by the (metric independent) standard relation
\bse\label{CartToSpherical}
\begin{align}
	x^1 &= r\sin \vartheta \cos \varphi,\\
	x^2 &= r\sin \vartheta \sin \varphi,\\
	x^3 &= r\cos \vartheta.
\end{align}
\ese
We will refer to all three spherical coordinates with the indices $a,b,\ldots$ while the angular coordinates will be referred to with $A,B,\ldots$. The angular coordinates will be collectively denoted as $\theta$ so that a single one of them will sometimes be denoted as $\theta^A$.

Using the map \eqref{exp} a neighborhood of $\sigma_0$ is covered with $q$-dependent Cartesian adapted coordinates $x^I$.  The components of a metric tensor $q$ with respect to the coordinates $x^I$ satisfy the following conditions at the point $\sigma_0$
\be\label{q0}
	q_{IJ}(\sigma_0) = \delta_{IJ},\qquad q_{IJ,K}(\sigma_0) = 0.
\ee
While, when expressed in the $q$-dependent spherical adapted coordinates $(r,\theta)$, the components of the metric tensor satisfy
\be\label{radialgauge}
	q_{rr} = 1,\qquad q_{rA} = 0.
\ee
Note, that those conditions are satisfied not only at $\sigma_0$ but in the entire neighborhood of $\sigma_0$ in which the spherical adapted coordinates are well-defined.

The space of the values of the $(x^I)$ coordinates, being also the space $\mathbb{R}^3$ from \eqref{exp}, is an internal space of the observer. She observes the world (that is $\Sigma$ and the canonical data defined on it) by pulling it back along the map $\xi$ to her internal space of labels. In particular, this space is endowed (globally) with a pullback of the metric tensor
\be
	Q = \xi^*q.
\ee
The resulting tensor $Q$ is symmetric, non-negative, locally near $(0,0,0)$ it is a metric tensor,
it   satisfies at every point of $\mathbb{R}^3$
\be\label{RAD}
	Q_{rr} = 1,\qquad Q_{rA} = 0,
\ee
and at $(0,0,0)$ 
\be
	Q_{IJ}(0,0,0) = \delta_{IJ},\qquad Q_{IJ,K}(0,0,0) = 0.
\ee
The conjugate momentum $p$ can be pulled back only locally, to define observer's $P$. Remarkably, an object globally defined on the observers internal space is 
\be
	\xi^* \left( p^{cd}\frac{q_{ca}q_{db}}{\sqrt{\det q}} \right).
\ee

We note here that two different metric tensors may define the same systems of adapted coordinates. This is the case if and only if the components of either of those metrics, when expressed in coordinates adapted to the other one, still satisfy conditions \eqref{radialgauge}. Of course the coordinates are constructed invoking the observer, so conditions \eqref{radialgauge} specify a gauge fixing for transformations preserving the observer. This is the radial gauge which was extensively studied in \cite{Bodendorfer:2015:RG1}. The purpose of the current paper is to study the transformations which do not preserve the observer.

%~~~~~~~~~~~~~~~~~~~~
\subsection{Spatial diffeomorphisms}

Observer's fields $Q, P$ are nothing else but the observables discussed extensively in \cite{Duch:2014hfa} and are, therefore, invariant with respect to the set of spatial diffeomorphisms
\be
	f : \Sigma\ \rightarrow\ \Sigma
\ee
which preserve the observer, i.e., such that  
\be
	f(\sigma_0) = \sigma_0 \qquad\text{and}\qquad e_I(f^*q) = (f^{-1})_*(e_I(q)) \text{ for }I=1,2,3,
\ee
where we have explicitly denoted the dependence of the frame on the spatial metric for clarity. These are observer's gauge diffeomorphisms, which we denote as Diff$_\text{obs}$. It is easy to check that Diff$_\text{obs}$ is a subgroup of all the spatial diffeomorphisms Diff.

%~~~~~~~~~~~~~~~~~~~~~
\subsection{Diffeomorphisms of the internal space}\label{sec:diff_internal_space}

The observer may consider diffeomorphisms of her internal space 
$\mathbb{R}^3$
\be
	F : \mathbb{R}^3\ \rightarrow\ \mathbb{R}^3.
\ee
Those diffeomorphisms may be viewed as active diffeomorphisms seen by the observer. However, a careful observer quickly notices that a generic diffeomorphism of her $\mathbb{R}^3$ destroys the condition \eqref{RAD}, and hence, it is not an interesting symmetry. Given a metric tensor $q$ on $\Sigma$ and its pullback $Q$ onto observer's $\mathbb{R}^3$, a symmetry is a map
\be
	F : \mathbb{R}^3\ \rightarrow\ \mathbb{R}^3
\ee
such that $F$ is an isomorphism in a neighborhood of $(0,0,0)$ and
\be
	F^*Q_{rr} = 1,\qquad F^*Q_{rA} = 0 .
\ee
Such observer's physical motions of her internal space correspond to the residual diffeomorphisms of $\Sigma$ complementary to the observer-preserving diffeomorphisms Diff$_\text{obs}$. 
 
For every metric tensor $q$, a diffeomorphism $f$ is called {\it residual} if it maps $q$ into $f^*q$ such that the coordinates adapted to $f^*q$ coincide in some neighborhood of the observer's location $\sigma_0$ with the coordinates adapted to $q$.  
In other words, Diff$_{\rm res}(q)$ are the diffeomorphisms preserving locally at $\sigma_0$, 
the radial gauge defined by the spherical coordinates adapted to $q$. Notice, that if $f\in \text{Diff}_\text{obs}\cap \text{Diff}_\text{res}(q)$, then, in a suitable neighborhood of $\sigma_0$,
\be
	f= \text{id}.
\ee 
Every residual diffeomorphism $f$ pulled back by \eqref{exp} defines in a neighborhood of $(0,0,0)$ an observer's physical  symmetry, 
\be
	F = \xi^{-1} \circ f \circ \xi.
\ee

Diff$_\text{obs}$ and the set of the residual diffeomorphisms Diff$_\text{res}(q)$ corresponding to a given observer and a metric $q$, generate all the diffeomorphisms Diff.

For clarity let us summarize the basic properties of the two classes of diffeomorphisms. Residual diffeomorphisms preserve the adapted coordinates and the gauge conditions \eqref{radialgauge}, but they do not preserve the observer (and they do not preserve the observables from \cite{Duch:2014hfa}). The Diff$_\text{obs}$ diffeomorphisms preserve the observer (and the observer's observables), but they do not preserve the radial gauge and the adapted coordinates.

As it will soon become clear, the residual diffeomorphisms Diff$_\text{res}(q)$ for a generic metric tensor $q$ do not form a subgroup of Diff, however, we will show that their induced action on the phase space $\Gamma$ does define a group. It follows from the fact, that the property of defining the same Cartesian (and spherical) adapted coordinates is an equivalence relation in the set of metric tensors.

%~~~~~~~~~~~~~~~~~~~~~~~~
\subsection{Induced action of the residual diffeomorphisms in $\Gamma$}\label{diffresdefinegroup3d}
%~~~~~~~~~~~~~~~~~~~~~~~~

Given an observer ${\cal O}$, an induced action of the residual diffeomorphisms in $\Gamma$ can be defined as follows.
Given a map on the space of metric tensors
\be\label{resact1}
	q\ \mapsto\  f_q\in \text{Diff}_\text{res}(q)
\ee
its action in $\Gamma$ is
\be\label{resact2}
	(q,p)\ \mapsto\ (f_q^*q,f_q^*p).
\ee
To see that such maps form a group, consider another map of this kind
\be
	q\ \mapsto\ \bar f_q\in \text{Diff}_\text{res}(q).
\ee
The composition
\be\label{comp}
	(q,p)\ \mapsto (f_q^*q, f_q^*p) \mapsto (\bar f_{f^*_q q}^*(f_q^*q), \bar f_{f^*_q q}^*(f_q^*p))
\ee
can be written as
\be\label{comp'}
	(q,p)\ \mapsto ((f_q\circ \bar f_{f^*_q q})^*q, (f_q\circ \bar f_{f^*_q q})^*p).
\ee 
Now, according to our assumptions
\be 
	x^I((f_q\circ \bar f_{f^*_q q})^*q) = x^I(\bar f_{f^*_q q}^*(f^*_q q)) = x^I(f^*_q q) = x^I(q),
\ee
where we have displayed the metric to which given coordinates are adapted for clarity and in the last two steps we have exploited the defining property of the residual diffeomorphisms. Therefore indeed, it follows that the composition is defined by a map
\be
	q\ \mapsto\  f_q\circ \bar f_{f^*_q q} \in \text{Diff}_\text{res}(q),
\ee
which completes the argument that the action of residual diffeomorphisms in the phase space defines a group, since only its closure was in question.

%~~~~~~~~~~~~~~~~~~~~~~~~
\subsection{Characterization of the residual diffeomorphisms}
%~~~~~~~~~~~~~~~~~~~~~~~~

To understand the action induced by the residual diffeomorphisms Diff$_\text{res}$ let us start by characterizing the residual diffeomorphisms themselves. Perhaps the simplest example of a residual diffeomorphism is specified by a rotation matrix, e.g.,
\be
	R: \mathbb{R}^3\rightarrow \mathbb{R}^3.
\ee
A diffeomorphism of $\Sigma$ is then described in coordinates adapted to $q$ by
\be
	x^I\ \mapsto\ R^I{}_Jx^J,
\ee
i.e., the diffeomorphism maps a point $\sigma\in\Sigma$ with coordinate values $x^I$ into a point with coordinates $R^I_Jx^J$.
This example may, therefore, be viewed as a rotation of the observer ${\cal O}$. Note however, that the rotation
depends on a metric tensor $q$, as it was specified in coordinates adapted to that metric.

A more general residual diffeomorphism consists of a rotation and a translation of the observer. Given a metric $q$, suppose that $f\in \text{Diff}_\text{res}(q)$, i.e., it preserves the radial gauge of $q$. Consider an auxiliary observer $\bar{\cal O}=(\bar\sigma_0,\bar e_I(\cdot))$ such that
\be
	\bar\sigma_0 = f(\sigma_0),\qquad \bar e_I =  f_*e_I.
\ee 
The Cartesian coordinates adapted to $q$ constructed for the auxiliary observer, denoted by $\bar x^I$, satisfy
\be
	f^*(\bar x^I(q)) = x^I(f^*q) = x^I(q),
\ee
where we have invoked explicitly to which metric given coordinates are adapted, the first equality follows from the covariance of the coordinates, and the second from the fact that $f$ is a residual diffeomorphism. Evaluation of the two sides at a point $\sigma\in\Sigma$ gives
\be\label{themap}
	\bar x^I(q; f(\sigma)) = x^I(q; \sigma),
\ee
where we have indicated both the metric to which coordinates are adapted and the points at which they are evaluated. In other words, the diffeomorphism is simply the active transformation defined by the two coordinate systems. The rotation example given above is a special case when $f(\sigma_0)=\sigma_0$, and the rotation matrix is then specified by
\be
	R = f_*(\sigma_0).
\ee

Conversely, if we are given a metric tensor $q$, then to construct $f\in \text{Diff}_\text{res}(q)$ we can choose any auxiliary observer $\bar{\cal O} = (\bar\sigma_0, \bar e_I(\cdot))$ and use her coordinates $\bar x^I$ adapted to $q$ and the formula \eqref{themap} to determine $f$.   

This indicates, that the family $\text{Diff}_\text{res}(q)$ of the residual diffeomorphisms is a 6-dimensional family of auxiliary observers parametrized by: $3$ rotations of $e_I(\cdot)$, and $3$ translations of $\sigma_0$. Below, we will find explicitly all the infinitesimal generators of Diff$_\text{res}(q)$.

%~~~~~~~~~~~~~~~~~~~~~~~~
\subsection{Generators of the action of Diff$_\text{res}$ in $\Gamma$}
%~~~~~~~~~~~~~~~~~~~~~~~~

In this section we will find all the infinitesimal generators of the action of residual diffeomorphisms spelled out in \eqref{resact2}. The generators are vector fields on $\Gamma$ of the form\footnote{The indices $i,j$ refer to some auxiliary coordinate system.}
\be\label{inf1}
	X =  \int d^3 \sigma \left( \delta q_{ij}(q;\sigma)\frac{\delta}{\delta q_{ij}(\sigma)} + 
\delta p^{ij}(q;\sigma)\frac{\delta}{\delta p^{ij}(\sigma)} \right).
\ee 
Since they generate the action of diffeomorphisms, their coefficients $\delta q_{ij}$ and $\delta p^{ij}$ are given by
\bse\label{inf2}
\begin{align}
	\delta q_{ij}(q;\sigma) = {\cal L}_{\Delta} q_{ij}(\sigma),\\
	\delta p^{ij}(q;\sigma) = {\cal L}_{\Delta} p^{ij}(\sigma),
\end{align}
\ese
where $\cal L$ is the Lie derivative and $\Delta$ is a vector field tangent to $\Sigma$ depending on $q$ (in general, vector fields on $\Gamma$ could also depend on the momentum $p$, but the conditions defining \eqref{resact2} did not involve $p$).
 
Finally, since the diffeomorphisms considered in the induced action belong to the corresponding Diff$_\text{res}(q)$, for every metric $q$, we have in terms of coordinates adapted to that metric
\be\label{inf3}
	{\cal L}_{\Delta} q_{rr} = 0,\qquad  {\cal L}_{\Delta} q_{rA} = 0.
\ee    
In conclusion, a vector field \eqref{inf1} is an infinitesimal generator of the action \eqref{resact2} of the residual  diffeomorphisms in $\Gamma$ if and only if it is given by \eqref{inf2} and \eqref{inf3}. 

\vspace{0.5cm}

The remaining task is to find explicit expressions for the vector field $\Delta$ and see that indeed it is fully specified by conditions \eqref{inf3}. Fortunately, this task has been completed already in \cite{Duch:2014hfa}, so  let us only recall the main steps here, for later reference. In the spherical adapted coordinates \eqref{inf3} take the form
\bse\label{1i2}
\begin{align}
	\nabla_r \Delta_r = 0,
	\\
	\nabla_r \Delta_{A} + \nabla_A \Delta_r = 0.
\end{align}
\ese
Exploiting the properties of the adapted coordinates noted in \eqref{radialgauge} we can rewrite those equations as
\bse\label{1oraz2}
\begin{align}
	\frac{\partial}{\partial r}\Delta^r(r,\theta) &= 0,\label{1}\\
	\frac{\partial}{\partial r}\Delta^A(r,\theta) &= -q^{AB}(r,\theta)\frac{\partial}{\partial \theta^B}\Delta^r(r,\theta),\label{2}
\end{align}
\ese
where $r=0$ corresponds to the point $\sigma_0$.  Of course, we are interested only in those solutions which define vector fields continuous and differentiable in $\sigma_0$. At this point, the first equation in \eqref{inf3} implies
\be
	{\cal L}_{\Delta} q_{ij}(\sigma_0) = 0.
\ee
In terms of the Cartesian adapted coordinates, this equation reads
\be\label{anty}
	\partial_J\Delta^I(\sigma_0) = - \partial_I \Delta^J(\sigma_0) .
\ee
Finally, the free initial data for the equations \eqref{1oraz2} is 
\be\label{in}
	\Delta^I(\sigma_0), \qquad\text{and}\qquad \partial_J\Delta^I(\sigma_0) \text{ subject to \eqref{anty}.}\ee

The complete set of solutions of \eqref{1oraz2} subject to \eqref{in} was derived in \cite{Duch:2014hfa}. There are 3 solutions $L_I$ corresponding to initial data such that $\Delta^I(\sigma_0) = 0$
and 3 solutions $ T_I$ corresponding to initial data such that $\Delta^I{}_{,J}(\sigma_0) = 0$. They read
\bse\label{LandT}
\begin{align}
	 L_I(x) &= L_I^J(x)\frac{\partial}{\partial x^J} = \epsilon_{IJK} \delta^{KL} x^J\frac{\partial}{\partial x^L}
	 = \epsilon_{IJ}{}^K x^J\frac{\partial}{\partial x^K},
	 \\
	 T_I(x) &= T_I^J(x)\frac{\partial}{\partial x^J} = \frac{\partial}{\partial x^I} -  \delta_{IJ} \, \int_0^1 \frac{d\lambda}{\lambda^2}\, \left( q^{JK}(\lambda x) - \delta^{JK}\right) \frac{\partial}{\partial x^K}.\label{T}
\end{align}
\ese

At this point we can conclude our task of characterizing the infinitesimal generators \eqref{inf1}.
Their general form is 
\bse\label{inf4}
\begin{gather}
	X^{(\Delta)} = \int d^3\sigma\left({\cal L}_{\Delta}q_{ij}(\sigma)\,\frac{\delta}{\delta q_{ij}(\sigma)} + {\cal L}_{\Delta}p^{ij}(\sigma)\,\frac{\delta}{\delta p^{ij}(\sigma)}\right)\\
\intertext{with}
	\Delta = t^I T_I + \omega^I L_I
\end{gather}
\ese
where $t^I$ and $\omega^I$ are 6 arbitrary coefficients arbitrarily depending on $q$. Notice, that the vector fields $L_I$ depend on $q$ via the adapted coordinates. So do $T_I$, and in addition, their coefficients also depend on $q$ explicitly.

%~~~~~~~~~~~~~~~~~~~~~~~~
\subsection{Commutators of the infinitesimal generators}
%~~~~~~~~~~~~~~~~~~~~~~~~

As was shown in subsection \ref{diffresdefinegroup3d} the residual diffeomorphisms form a group acting on the phase space, thus their generators $X^{(\Delta)}$ form a Lie algebra, i.e., a commutator of two of them, say $X^{(\Delta)}$ and $X^{({\bar\Delta})}$, is given by
\be\label{alg}
	[X^{(\Delta)}, X^{({\bar\Delta})}] = X^{({\bar{\bar\Delta}})},
\ee
where the resulting $X^{({\bar{\bar\Delta}})}$ is also given by \eqref{inf4} with some ${\bar{\bar\Delta}}$. To see the dependence of $\bar{\bar \Delta}$ on $\Delta$ and $\bar\Delta$ we use the formula \eqref{inf4}
\begin{multline}\label{generalformofcomm}
	[X^{(\Delta)}, X{}^{(\bar{\Delta})}] = X^{(\Delta)}(X^{(\bar{\Delta})})-X^{(\bar{\Delta})}(X^{(\Delta)})
	\\
	= X^{(\Delta)}\left(\int d^3\sigma
	\Big({\cal L}_{\bar{\Delta}}q_{ij}\frac{\delta}{\delta q_{ij}}+{\cal L}_{\bar{\Delta}}p^{ij}\frac{\delta}{\delta p^{ij}}\Big)
	\right) 
	- (\Delta \leftrightarrow \bar\Delta)
	%- X^{(\bar{\Delta})}\left(\int d^3\sigma\big({\cal L}_{\Delta}q_{ij}\frac{\delta}{\delta q_{ij}} + {\cal L}_{\Delta}p^{ij}\frac{\delta}{\delta p^{ij}}\big)\right)
	\\
	= \int d^3\sigma\Big(
	{\cal L}_{\bar{\Delta}}{\cal L}_{\Delta}q_{ij}\frac{\delta}{\delta q_{ij}}
	+{\cal L}_{\bar{\Delta}}{\cal L}_{\Delta}p^{ij}\frac{\delta}{\delta p^{ij}}\Big) 
	+ \int d^3\sigma\Big({\cal L}_{X^{(\Delta)}(\bar{\Delta})}q_{ij}\frac{\delta}{\delta q_{ij}}+{\cal L}_{X^{(\Delta)}(\bar{\Delta})}p^{ij}\frac{\delta}{\delta p^{ij}}\Big)
	- (\Delta \leftrightarrow \bar\Delta)
	%
	%- \int d^3\sigma\big({\cal L}_{\Delta}{\cal L}_{\bar{\Delta}}q_{ij}\frac{\delta}{\delta q_{ij}} + {\cal L}_{\Delta}{\cal L}_{\bar{\Delta}}p^{ij}\frac{\delta}{\delta p^{ij}}\big)	
        % 
        %- \int d^3\sigma\big({\cal L}_{X^{(\bar{\Delta})}(\Delta)}q_{ij}\frac{\delta}{\delta q_{ij}} + {\cal L}_{X^{(\bar{\Delta})}(\Delta)}p^{ij}\frac{\delta}{\delta p^{ij}}\big)
	\\
	= X^{(-[\Delta,\bar{\Delta}] + X^{(\Delta)}(\bar{\Delta}) - X^{(\bar{\Delta})}(\Delta))},
\end{multline}
where $X^{(\Delta)}(\bar\Delta)$ denotes the vector field on $\Sigma$ resulting from the action of the infinitesimal generator $X^{(\Delta)}$ on the vector field $\bar\Delta$ treated as a functional on the phase space $\Gamma$. In this way we see that
\be
	{\bar{\bar\Delta}} = -[\Delta,\bar\Delta] + X^{(\Delta)}(\bar\Delta) -  X^{(\bar\Delta)}(\Delta),
\ee

Although in principle the last two terms determining the field $\bar{\bar\Delta}$ in \eqref{generalformofcomm} are very complicated, there are a number of simplifications present due to the peculiarities of the considered construction. The infinitesimal generators act as Lie derivatives on the phase space variables along the fields being their parameters. Since we consider vector fields generating residual diffeomorphisms, i.e. preserving the adapted coordinates, the only phase-space dependence which will be seen by those generators is the explicit dependence. Note however, that only the generators of translations (spelled out in \eqref{T}) depend explicitly on the phase space, and moreover, they depend only on the spatial metric. This will vastly simplify the derivation of the commutation relations for the particular base vector fields from \eqref{LandT}.

Let us analyze them in the order of growing complexity.

%~~~~~~~~~~~~~~~
\subsubsection{$ [X^{(L_I)}, X^{(L_J)}]$}\label{sec_commutator_LL}

The simplest to establish is the commutator of two generators of rotations. As we have stated in the previous section, even though the vector fields $L_J$ depend on $q$ via the adapted coordinates, since the rotations belong to the residual diffeomorphisms Diff$_\text{res}$, they preserve the adapted coordinates. Therefore,
\be
	[X^{(L_I)}, X^{(L_J)}] = X^{(-[L_I,L_J])} = \epsilon_{IJ}{}^K X^{(L_K)},
\ee
where the last step is a straightforward calculation.

%~~~~~~~~~~~~~~~
\subsubsection{$[X^{(L_I)}, X^{(T_J)}]$}\label{sec_commutator_LT}

The second easiest is the commutator of a generator of a rotation with a generator of a translation. What changes now is that the coefficients of the vector field $T_J$ depend on the metric tensor $q$. Therefore, the general formula \eqref{generalformofcomm} becomes
\be
	[X^{(L_I)}, X^{(T_J)}] = X^{(-[L_I,T_J] + X^{(L_I)}(T_J))},
\ee
where, more specifically, $X^{(L_I)}(T_J)$ denotes the vector field on $\Sigma$ whose value at a point $\sigma\in\Sigma$ corresponding to the values $x^I$ of the coordinates adapted to $q$ is   
\be
	X^{(L_I)}(T_J)(x) =  X^{(L_I)}(T_J^K(x))\frac{\partial}{\partial x^K},
\ee
where $T_J^K(x)$ at fixed $x^I$ is considered as a function on the phase space $\Gamma$. 

It is easy to understand the meaning of the difference $-[L_I,T_J] + X^{(L_I)}(T_J)$. The first term describes a rotation of all of $T_J$ and a differentiation. The second term describes a compensating contribution coming from a rotation of the factors of $q^{JK}(\lambda x)$. By looking at the form of $T_J$ it is possible to see, that 
\be
	-[L_I,T_J] + X^{(L_I)}(T_J) = \epsilon_{IJ}{}^KT_K
\ee  
Hence,
\be
	[X^{(L_I)}, X^{(T_J)}] = \epsilon_{IJ}{}^K X^{(T_K)}.
\ee

%~~~~~~~~~~~~~
\subsubsection{$[X^{(T_I)}, X^{(T_J)}]$}

It remains to compute the commutator of two generators of translations. Several general properties of our vector fields will let us determine the result. As before, we have the general formula
\be\label{translationscommutatorgeneral}
	[X^{(T_I)}, X^{(T_J)}] = X^{(-[T_I,T_J] + X^{(T_I)}(T_J) - X^{(T_J)}(T_I))}.
\ee
On the other hand, we conclude from (\ref{alg}), that for every given metric tensor $q$ it holds that
\be\label{TT}
    -[T_I,T_J] + X^{(T_I)}(T_J) - X^{(T_J)}(T_I) = t^M T_M + \omega^M L_M
\ee
for some (possibly metric dependent) coefficients $t^I$ and $\omega^I$. To identify them we shall compute the values at $\sigma_0$ of the left hand side of the above equation and its first derivative and compare the result with the data for the generators $L_I$ and $T_I$ which is
\bse\label{RcommaK}
\begin{align}
T_M^K(\sigma_0) &= \delta^K_M,\qquad  L_M^K(\sigma_0) = 0,  \\
T_{M,L}^K(\sigma_0) &= 0, \qquad L_{M,L}^K(\sigma_0) = \epsilon_{ML}{}^K.
\end{align}
\ese
From the form of the generators of translations spelled out in \eqref{T} we see that at the location of the observer
\be
	T_{I,MN}^K(\sigma_0) = - \delta_{IJ}\int_0^1 d\lambda\, (\partial_M \partial_N q^{JK})(\sigma_0) = - q_{IJ}(\sigma_0) \partial_M \partial_N q^{JK}(\sigma_0) = q^{JK}(\sigma_0)q_{IJ,MN}(\sigma_0).
\ee
Thus, $T_I^J(x)$ has the following Taylor expansion around $\sigma_0$
\be
	T_I^K(x) = \delta_I^K + \frac{1}{2}q^{JK}(\sigma_0)q_{IJ,MN}(\sigma_0) x^M x^N + O(r^3).
\ee 
It follows, that both $X^{(T_I)}(T_J)$ and $X^{(T_J)}(T_I)$ are of the order $r^2$, while $[T_I,T_J]$ is of the order of $r$. Hence, we can conclude that
\be
	\left( -[T_I,T_J] + X^{(T_I)}(T_J) - X^{(T_J)}(T_I) \right)(\sigma_0) = 0,
\ee
and
\begin{multline}\label{[TT]}
	\partial_L\left( -[T_I,T_J] + X^{(T_I)}(T_J) - X^{(T_J)}(T_I) \right)^K(\sigma_0) 
	= \partial_L\left( -[T_I,T_J]  \right)^K(\sigma_0) 
	\\
	= T_{I,JL}^K(\sigma_0) - T_{J,IL}^K(\sigma_0)
	= q^{MK}(\sigma_0) (q_{IM,JL}(\sigma_0) - q_{JM,IL}(\sigma_0))
	=- R^K{}_{LIJ}(\sigma_0),
\end{multline}
where in the last step we applied the identity \ref{eq:app_riemann_norm_coor} from Appendix \ref{app:prop_norm_coor}. It follows from equations \eqref{TT}, \eqref{RcommaK} that $t^M \delta^K_M = 0$ and $\omega^M\epsilon_{ML}{}^K = - R^K{}_{LIJ}(\sigma_0)$. Therefore,
\be
	[X^{(T_I)}, X^{(T_J)}] = \frac{1}{2}R_{IJLM}(\sigma_0)\epsilon^{LMK}X^{(L_K)}.
\ee

Note that to obtain the above result the knowledge of the exact form of the fields $T_I$ and $L_I$ everywhere was not necessary -- we only had to know the values at $\sigma_0$ of $T_I$, $L_I$ and their derivatives up to, at most, second order. We could have used a similar method to compute the commutators involving one or two generators of rotations, instead of performing the explicit calculations which we presented in the subsections \ref{sec_commutator_LL} and \ref{sec_commutator_LT}. In fact, in the spacetime case, discussed in section \ref{spacetimecase}, we will not use the explicit form of the generators at all, deriving the full algebra invoking only their values and values of their derivatives (up to second order) at the location of the observer.

The above result concludes our investigation of the algebra of the generators of the action of the residual diffeomorphisms Diff$_\text{res}$ in the phase space $\Gamma$.

%~~~~~~~~~~~~~~~~~~~
\subsection{Resulting algebra}\label{sec:ResultingAlgebra3d}

Collecting the results from the previous sections we can say that if
\be
	\Delta = t^I T_I + \omega^I L_I \quad\text{and}\quad \bar\Delta = \bar t^I T_I + \bar \omega^I L_I,
\ee
then $\bar{\bar\Delta} = \bar{\bar t}^I T_I + \bar{\bar \omega}^I L_I$ such that
\be
	[X^{(\Delta)},X^{(\bar\Delta)}] = X^{(\bar{\bar\Delta})},
\ee
is parametrized by
\bse\label{algebraspatial}
\begin{align}
	\bar{\bar t}^I &= t^J \bar\omega^K \epsilon_{JK}{}^I - \bar t^J\omega^K\epsilon_{JK}{}^I,\\
	\bar{\bar \omega}^I &= \frac{1}{2} t^J \bar t^K R_{JKLM}(\sigma_0)\epsilon^{LMI} + \omega^J\bar\omega^K\epsilon_{JK}{}^I.
\end{align}
\ese

%~~~~~~~~~~~~~~~~~~~
\section{Residual diffeomorphisms in the spacetime case}\label{spacetimecase}
%~~~~~~~~~~~~~~~~~~~

The previous section was a preparation for a more advanced case of a spacetime observer, which we consider in this section. As before, we will study the residual diffeomorphisms seen by the observer and derive the algebra of infinitesimal generators they induce in the space of spacetime metric tensors.\footnote{This derivation has been a part of  \cite{DuchMagisterka}.}

In this section we will use a rather general derivation of the result, in which the explicit form of the generators will not be relevant, contrary to our strategy in the previous section. We note here, that we could have used a similar reasoning also in the spatial case.

%~~~~~~~~~~~~~~~~~~~~
\subsection{Spacetime observer}\label{sec:defObserver4d}

The definition of an observer in the spacetime case is analogous to that in the spatial case (cf.~section \ref{sec:defObserver3d}). To specify a spacetime observer we need to fix 
\begin{enumerate}[label=\it{(\roman*)}]
\item a point $p$ in spacetime $\mathcal{M}$ representing the location of an observer and
\item a map assigning an orthonormal tangent frame at $p$ to each metric tensor $g$ on $\mathcal{M}$
\be
	g\ \mapsto\ e_\mu\in T_{p}\mathcal{M},
\ee
where $\mu\in\{0,1,2,3\}$ is a label and by definition the vector $e_0\equiv e_\tau$ is timelike (it has the interpretation of the initial four-velocity of the observer). 
\end{enumerate}

%~~~~~~~~~~~~~~~~~~~~
\subsection{Spacetime observer's internal space of labels}

As in the spatial case the observer probes the spacetime $\mathcal{M}$ using certain metric dependent coordinates, which we again call the adapted coordinates\footnote{In the literature they are also known as Fermi coordinates.}. To define them we first specify a spacetime geodesic $\gamma(\tau)$ representing the worldline of an observer which is such that
\be
	\gamma(0)=p,\qquad \left.\frac{\text{d}}{\text{d}\tau}\right|_{\tau=0}\gamma = e_\tau.
\ee
Next, we define a family of spatial orthonormal frames $E_I$ ($I\in\{1,2,3\}$) which are parallel transports of $e_I$  along the worldline of the observer. We complete the definition of the adapted coordinates by using the exponent map and the frames $E_I(\tau)$ for all $\tau$'s to cover (a tubular neighborhood of $\gamma$ in) the spacetime around the observer. We denote this parametrization by $\xi[p,e_\mu]$
\be\label{eq:defCoordiantes4d}
	\xi[p,e_\mu]:\mathbb{R}^4\ni(\tau,x^I) \mapsto \exp_{\gamma(\tau)}(x^I E_I(\tau))\in{\cal M}.
\ee
In this way on the worldline of the observer we have
\be
	 x^I(\gamma(\tau)) = 0.
\ee
Note that the map $\xi[p,e_\mu]$ is injective only in a tubular neighborhood of the time coordinate axis, due to standard problems with formation of caustics of geodesics. We will refer to the coordinates $(\tau,x^I)$ with indices $\mu,\nu,\ldots$. We will also invoke their spherical partners, which differ from the ones defined above by the standard change \eqref{CartToSpherical} performed on the three spatial coordinates of the above mentioned ones.

The components of the metric tensor in the adapted coordinates we have just introduced satisfy the following conditions along the worldline of the observer (compare with \eqref{q0})
\be\label{g0}
	g_{\mu\nu} = \eta_{\mu\nu}, \qquad \partial_\mu g_{\nu\rho} = 0,
\ee
where $\eta_{\mu\nu}$ is the Minkowski metric. Moreover, we have (compare with \eqref{radialgauge})
\be\label{radialgauge4d}
		g_{r\tau} = 0,\qquad g_{rr} = 1,\qquad g_{rA} = 0
\ee
in the spherical adapted coordinates in a tubular neighborhood of $\gamma$ in which the coordinates are well-defined.

When comparing the definitions of the observer and her adapted coordinates in the spatial (discussed in section \ref{spatialcase}) and spacetime cases the close analogy is apparent. The key difference of the current case with respect to the spatial case is that now we use an exponent map which sends spacetime geodesics and not spacelike geodesics as the map \eqref{exp} did. This is reflected in the fact that the coordinate lines of the radial adapted coordinate were spatial geodesics (i.e., geodesics of the spatial 3-metric) in the spatial case, but are spacetime geodesics (i.e., geodesics of the spacetime 4-metric) in the current case. When viewing the construction from the gauge-fixing point of view, this can be translated into one extra gauge condition which needs to be added to complete the conditions \eqref{radialgauge} (see \cite{Bodendorfer:2015kua} for details). Hence the spacetime radial gauge is specified by the conditions
\be\label{spacetimeRG}
		q_{rr} = 1,\qquad q_{rA} = 0, \qquad K_{rr} = 0,
\ee
where $K$ is the extrinsic curvature tensor of a spatial slice, being a simple function of the canonical momentum and $q$ is the spatial metric induced from the spacetime metric $g$ on that spatial slice. As underlined in \cite{Bodendorfer:2015kua}, the key source of problems with the spacetime radial gauge is that the gauge conditions do not mutually weakly Poisson commute (since the extra condition $K_{rr}=0$ does not Poisson commute with condition $q_{rr}=1$), which has a direct consequence in the form of the algebra of the Dirac bracket. However leaving those problems aside, let us derive a result analogous to the one presented in section \ref{spatialcase}.

%~~~~~~~~~~~~~~~~~~~~
\subsection{Residual diffeomorphisms -- transformations between observers}

Given a metric tensor $g$, a residual diffeomorphism is, by definition, a transformation $f:\mathcal{M}\rightarrow\mathcal{M}$ such that the coordinates adapted to $f^*g$ coincide in some neighborhood of the observer's worldline $\gamma$ with the coordinates adapted to $g$ (cf.~the definition of the residual diffeomorphism in the spatial case given in section \ref{sec:diff_internal_space}). It means that Diff$_{\rm res}(g)$ are the diffeomorphisms preserving locally the spacetime radial gauge defined in the previous section.

We will show that the residual diffeomorphisms arise naturally as transformations between observers. To this end, beside the fixed observer $\mathcal{O} = (p,e_\mu)$ let us consider another one $\bar{\mathcal{O}}= (\bar p,\bar e_\mu)$. We can define the following map
\be\label{eq:residualTransf4d}
	f_g = \xi[\bar p,\bar e_\mu]\circ\xi[p,e_\mu]^{-1},
\ee
where $\xi[p,e_\mu]$ is a parametrization \eqref{eq:defCoordiantes4d} associated to the observer ${\cal O} = (p,e_\mu)$. Since the map $f$ sends the point $\xi[p,e_\mu](\tau,x^I)\in\mathcal{M}$ into $\xi[\bar p,\bar e_\mu](\tau,x^I)\in\mathcal{M}$ it preserves the conditions \eqref{g0} and \eqref{radialgauge4d} fulfilled by components of the metric written in the adapted coordinates. Thus, the map \eqref{eq:residualTransf4d} is a residual transformation. It turns out that all residual transformations are of this form.

Like in the spatial case, given a generic spacetime metric tensor $g$, the set Diff$_{\rm res}(g)$ of the corresponding residual diffeomorphisms is not a group. The residual diffeomorphisms  induce a group of motions of the space $G$ of all the metric tensors defined on the manifold $\cal M$.\footnote{Note, that we could easily enlarge this space, e.g., by considering the space of all pairs consisting of a metric tensor and some other tensor field of a given type.} The group is formed by the action on $G$ of metric dependent residual diffeomorphisms 
\be
	G \ni g \mapsto f_g\in {\rm Diff}_{\rm res}(g) ,
\ee
as in the spatial case discussed in detail in section \ref{diffresdefinegroup3d}.

%~~~~~~~~~~~~~~~~~~~
\subsection{Generators of residual transformations}

Now, let us find the generators of the residual transformations. Given a metric tensor $g$, we first choose a one parameter family of observers
\be
	s \mapsto {\cal O}_s = (p(s),e_\mu(s)),
\ee
such that
\be
	p(0) = p, \qquad e_\mu(0) = e_\mu, \qquad g_{p(s)}(e_\mu(s),e_\nu(s)) = \eta_{\mu\nu},
\ee
where $g_{p(s)}$ is the spacetime metric at the point $p(s)$ and $\eta_{\mu\nu}$ is the Minkowski metric. Since the frame $e_\mu$ is necessarily orthogonal, its variations are limited to be of the form
\be
	\delta e_\mu = l_\mu{}^\nu e_\nu,
\ee
where $l_\mu{}^\nu$ is a generator of Lorentz transformations belonging to the defining representation of $SO(1,3)$. We parametrize variations of the location of the observer with
\be
	\delta p = t^\mu e_\mu.
\ee

The above family of observers defines a one parameter family of residual transformations
\be\label{eq:residualTransf4dgroup}
	f_g(s) = \xi[p(s),e_\mu(s)]\circ\xi[p,e_\mu]^{-1}.
\ee
Let us consider the vector field $Z$ defined as follows
\be\label{definicjaZ}
	Z[p,e_\mu;\delta p,\delta e_\mu]  = 
	\left.\frac{\text{d}}{\text{d} s}\right|_{s=0}\xi[p(s),e_\mu(s)]\circ\xi[p,e_\mu]^{-1},
\ee
Since the vector fields $Z[p,e_\mu;\delta p,\delta e_\mu]$ are infinitesimal residual transformations they satisfy the following equations  
\bse\label{eq:determineZ}
\begin{align}
	\eta_{\mu\rho} \partial_\nu Z^\rho + \eta_{\rho\nu} \partial_\mu Z^\rho &= 0, \label{eq:determineZfirst}
	\\
	\partial_\nu\partial_\mu Z^\rho + Z^\sigma \partial_\sigma\Gamma^\rho_{\phantom{\rho}\mu\nu} &=0,\label{eq:determineZsecond}
	\\
	{\cal L}_Z g_{r\mu} &= 0,\label{eq:determineZthird}
\end{align}
\ese
where the first two equations hold on the observer's worldline (in the adapted coordinates) whereas the last in the neighbourhood of this line (in the spherical adapted coordinates). The above identities were proven in Appendix \ref{appendixdrugapochodnaZ}. They follow from the fact, that residual transformations preserve the conditions \eqref{g0} and \eqref{radialgauge4d}. The equation \eqref{eq:determineZthird} might be written equivalently as
\be\label{eqNabla4d}
	\nabla_r Z^\mu + g^{\mu\nu}\nabla_\nu Z^r = 0,
\ee
which is a generalization of equations \eqref{1i2}. The set of equations \eqref{eq:determineZfirst}--\eqref{eq:determineZthird} yields a unique solution when supplemented with the initial conditions
\bse\label{initialconditionsforZ}
\begin{align}
	Z^\mu(p) &= t^\mu,\\
	\partial_\mu Z^\nu(p) &= l_\mu{}^\nu
\end{align}
\ese
(see the last paragraph of Appendix \ref{ZinDiracmatrix}), thus, the ten numbers $t^\mu$, $l_\mu{}^\nu$ parametrize the space of all possible generators of the residual diffeomorphisms we are considering.

The vector fields $Z$ discussed here, play an important role also when the gauge-fixing perspective is concerned, i.e., in the spacetime radial gauge. As shown in Appendix \ref{ZinDiracmatrix}, they belong to the kernel of the transposition of a part of the Dirac matrix in the spacetime radial gauge, discussed in \cite{Bodendorfer:2015kua}.

Since we are interested in the action of residual transformations in the space of Lorentzian metric tensors, their generators have the form
\be
 X^{(Z)} = \int d^4x \,{\cal L}_Z g_{\mu\nu}(x) \frac{\delta}{\delta g_{\mu\nu}(x)},
\ee
where $Z$ is the vector field defined above (cf.~equation \eqref{inf4}).

%~~~~~~~~~~~~~~~~~~~~
\subsection{Algebra of generators of residual transformations}

In order to find the algebra of transformations between observers we will consider a generator $X^{(\bar{\bar Z})}$ parametrized by $\bar{\bar t}^\mu$ and $\bar{\bar l}_\mu{}^\nu$ defined as a commutator of generators $X^{(Z)}$ and $X^{(\bar Z)}$, parametrized by $t^\mu$, $l_\mu{}^\nu$ and $\bar t^\mu$, $\bar l_\mu{}^\nu$, respectively. When we translate this condition to the vector fields we get (for comparison see \eqref{generalformofcomm})
\be\label{Zbarbar}
	\bar{\bar Z}[g] = - [Z[g], \bar Z[g]] + \delta\bar Z[g;{\cal L}_Z g] - \delta Z[g;{\cal L}_{\bar Z}g],
\ee
where we have noted the functional dependence on the metric explicitly and where we used the notation
\be
	\delta Z[g;\delta g] := \left.\frac{\text{d}}{\text{d}s}\right|_{s=0}Z[g+s\delta g].
\ee
Because the residual transformations form a group, the vector field \eqref{Zbarbar} we have just defined, is again of the form \eqref{definicjaZ}. 

Like in the spatial case, to obtain the structure constants (or functions) of the algebra, it is enough to analyze the field $\bar{\bar Z}$ (and its derivative) at the location of the observer, that is at point $p$. Since the initial conditions \eqref{initialconditionsforZ} do not depend on the metric we see that
\bse
\begin{align}
	\delta\bar Z^\mu[g;{\cal L}_Z g](p) &= 0,\qquad \delta Z^\mu[g;{\cal L}_{\bar Z}g](p) = 0,\\
	\partial_\mu\delta\bar Z^\nu[g;{\cal L}_Z g](p) &= 0,\qquad \partial_\mu\delta Z^\nu[g;{\cal L}_{\bar Z}g](p) = 0,
\end{align}
\ese
which means that in fact, only the first term on the right-hand side of \eqref{Zbarbar} will be important for our analysis. Therefore, we get
\be
	\bar{\bar t}^\mu = \bar{\bar Z}^\mu(p) = - [Z,\bar Z]^\mu(p) = - t^\nu \partial_\nu \bar Z^\mu(p) + \bar t^\nu \partial_\nu Z^\mu(p) = - t^\nu \bar l_\nu{}^\mu + \bar t^\nu l_\nu{}^\mu
\ee
and 
\begin{align}\label{obliczenieL}
	\bar{\bar l}_\mu{}^\nu &= \partial_\mu\bar{\bar Z}^\nu(p) = - \partial_\mu[Z,\bar Z]^\nu(p) \nonumber\\
	 &= - t^\rho(\partial_\mu\partial_\rho \bar Z^\nu)(p) + \bar t^\rho(\partial_\mu\partial_\rho Z^\nu)(p) - (\partial_\mu Z^\rho)(p)(\partial_\rho \bar Z^\nu)(p) + (\partial_\mu \bar Z^\rho)(p)(\partial_\rho Z^\nu)(p).
\end{align}
Taking into account the identity
\be\label{drugaprzezwartosc}
	\partial_\mu\partial_\rho Z^\nu (p) =  - Z^\sigma(p)(\partial_\sigma\Gamma^\nu{}_{\mu\rho})(p),
\ee
proven in Appendix \ref{appendixdrugapochodnaZ}, we finally obtain  
\be
\bar{\bar l}_\mu{}^\nu= t^\rho\bar t^\sigma(\partial_\sigma\Gamma^\nu{}_{\mu\rho})(p) - \bar t^\rho t^\sigma(\partial_\sigma\Gamma^\nu{}_{\mu\rho})(p) - l_\mu{}^\rho \bar l_\rho{}^\nu + \bar l_\mu{}^\rho l_\rho{}^\nu \nonumber\\
	 = t^\rho \bar t^\sigma R_{\rho\sigma\mu}{}^\nu(p) - l_\mu{}^\rho \bar l_\rho{}^\nu + \bar l_\mu{}^\rho l_\rho{}^\nu.
\ee
In the last step we used the fact that in the location of the observer in her adapted coordinates
\be\label{gammaznika}
	\Gamma^\nu{}_{\mu\rho}(p) = 0,
\ee
which follows directly from the second property in \eqref{g0} (in fact, the above equation holds along the entire worldline of the observer).

%~~~~~~~~~~~~~~~~~~
\subsection{Resulting algebra}\label{sec:ResultingAlgebra4d}

Collecting the results from the previous section, we can say that if $Z$ and $\bar Z$ are parametrized by $t^\mu,l_\mu{}^\nu$ and $\bar t^\mu,\bar l_\mu{}^\nu$ respectively, then $\bar{\bar Z}$ such that
\be
	[X^{(Z)},X^{(\bar Z)}] = X^{(\bar{\bar Z})},
\ee
is parametrized by
\bse\label{algebraspacetime}
\begin{align}
	\bar{\bar t}^\mu &= - t^\nu\bar l_\nu{}^\mu + \bar t^\nu l_\nu{}^\mu,\\
	\bar{\bar l}_\mu{}^\nu &= t^\rho \bar t^\sigma R_{\rho\sigma\mu}{}^\nu(p) - l_\mu{}^\rho \bar l_\rho{}^\nu + \bar l_\mu{}^\rho l_\rho{}^\nu.
\end{align}
\ese
This result is analogous to that obtained in the spatial case (cf.~\eqref{algebraspatial}) when the translations are required to be spatial and the only allowed Lorentz transformations are the rotations, namely
\be
	l_{IJ} = \epsilon_{IJK}\omega^K.
\ee

%~~~~~~~~~~~~~~~~~~~
\section{Comments}\label{sectionComments}
%~~~~~~~~~~~~~~~~~~~

In this section we collect several remarks concerning the presented results.
\begin{itemize}

\item An alternative, and in some sense closer to the spatial case, way of generalizing the spatial construction to the spacetime case would be to use spacetime Riemann normal coordinates. Having the frame at the location of the observer, we would then specify the coordinates in her spacetime neighborhood by labelling them with their spacetime distance from the observer.

The key problem of such a construction is due to the fact that while in the spatial case we parametrize a space equipped with metrics of euclidean signature in this way, in the spacetime case the signature is Lorentzian. This means that the surfaces of constant radial distance would become families of hyperboloidal surfaces. Moreover, the entire future and past lightcones with tips at the location of the observer, would acquire a zero value of the distance parameter. In contrast, using the coordinates we have described in section \ref{spacetimecase}, sometimes called Fermi coordinates, we have a physically natural construction at hand.

\item It is clear that the sets Diff$_\text{res}$(q) and Diff$_\text{res}$(g) are not groups for a generic Riemannian manifold $(\Sigma,q)$ and a generic pseudo-Riemannian manifold $({\cal M},g)$, respectively. In both spatial and spacetime cases, there are however three important exceptions when the set of residual transformations is a group, as summarized in the table
\begin{center}
\begin{tabular}{c|c}
 $(\Sigma,q)$ & Diff$_\text{res}$(q)
 \\
 \hline
 sphere with standard metric & $SO(4)$
 \\
 Euclidean space & Euclidean group $ISO(3)$
 \\
 hyperboloid with standard metric & Lorentz group $SO(1,3)$
\end{tabular}
\end{center}
for the spatial case, and for the spacetime case
\begin{center}
\begin{tabular}{c|c}
 $({\cal M},g)$ & Diff$_\text{res}$(g)
 \\
 \hline
 de Sitter spacetime & $SO(1,4)$
 \\
 Minkowski spacetime & Poincar\'e group $ISO(1,3)$
 \\
 anti-de Sitter spacetime  & $SO(2,3)$
\end{tabular}
\end{center}

In each of the above cases the algebra of generators of residual transformations (as given in section \ref{sec:ResultingAlgebra3d} or \ref{sec:ResultingAlgebra4d}) is equivalent to the Lie algebra of the appropriate group.

\item To define the observer in both the spatial and spacetime cases, in particular, one has to fix for each given metric the orthonormal frame at the location of the observer. This issue has been carefully discussed in \cite{Duch:2014hfa, Bodendorfer:2015:RG1} where to this end the Gram-Schmidt orthonormalization procedure was used.
%, forcing the initial conditions of equations defining the generators of diffeomorphisms preserving the observer Diff$_\text{obs}$ to be lower triangular.

However, in the current paper we were only invoking the residual diffeomorphisms Diff$_\text{res}$ and it turns out that the structure of the algebras of the generators of those diffeomorphisms, presented in \eqref{algebraspatial} and \eqref{algebraspacetime}, does not depend on the details of the construction of the orthonormal frame. Let us check this explicitly in the spacetime case. If, apart from the (metric dependent) choice of orthonormal frame $e_\mu$, we have an alternatively constructed (metric dependent) choice of orthonormal frame $\bar e_\mu$ than they are necessarily linked by a (metric dependent) Lorentz transformation
\be
	\bar e_\mu = L_\mu{}^\nu e_\nu.
\ee
The sets of parameters of generators of the residual diffeomorphisms are then linked by
\be
	\bar t^\mu = L_\nu{}^\mu t^\nu, \qquad \bar l_\nu{}^\mu = L_\rho{}^\mu L_\nu{}^\sigma l_\sigma{}^\rho.
\ee
Since $L_\nu{}^\mu$ depend only on the metric at the location of observer and the residual diffeomorphisms preserve the form of the metric at this point $\delta L_\nu{}^\mu[g;{\cal L}_Z g] = 0$ for any generator $Z$ and one may treat $L_\nu{}^\mu$ as independent of the metric $g$ in the computation of the commutator of the generators associated with the frame $\bar{e}_\mu$. Thus, the fact that $t$ and $l$ satisfy relations \eqref{algebraspacetime} for
\be
	R^\sigma{}_{\rho\mu\nu} e_\sigma = R(e_\mu, e_\nu) e_\rho,
\ee
implies that $\bar t$ and $\bar l$ satisfy those relations for
\be
	\bar R^\sigma{}_{\rho\mu\nu} \bar e_\sigma = R(\bar e_\mu, \bar e_\nu) \bar e_\rho,
\ee
where $R(\cdot, \cdot)$ is the Riemann tensor. This clarifies in what sense the structure of the algebra does not depend on the choice of an orthonormal frame in the definition of the observer.

\end{itemize}

%~~~~~~~~~~~~~~~~~~~
\section{Summary}
%~~~~~~~~~~~~~~~~~~~

We have investigated the residual diffeomorphisms present in the context of the observer's observables and the radial gauge. Although, given a generic metric tensor, they do not form a group due to lack of closure, their induced action on the phase space always closes. This situation is familiar from the Dirac algebra, i.e., the algebra of the constraints of canonical general relativity. There, one also faces structure constants (sometimes called functions) which depend on the point in the phase space.

In the spatial case, reviewed in section \ref{spatialcase}, we obtained explicit forms of the generators tangent to the phase space, and presented a derivation of their algebra. The generating vector fields have a clear interpretation of being translations and rotations of the observer, due to them being defined by their action at the location of the observer. This interpretation is strengthened by the fact that in flat space the algebra of the generators is just the Euclidean algebra (see the second comment in section \ref{sectionComments}). In general, however, the commutator of two translations does not vanish as there is an additional term proportional to the curvature of the spatial slice. The modification is present, since the translations are curved-space (metric-dependent) generalizations of flat-space translations.

The residual diffeomorphisms present in the spacetime version of the construction of observables, being the other face of the spacetime radial gauge, were analyzed in section \ref{spacetimecase}. Also here, the diffeomorphisms can be interpreted as translating and Lorentz-rotating the observer. The generators of the action of those diffeomorphisms on the space of metrics  form an algebra which for Minkowski spacetime is the Poincar\'e algebra (see the second comment in section \ref{sectionComments}). However, in general a modification proportional to the Riemann tensor is again present, because in a curved spacetime the commutator of translations probes its curvature.

An interesting consequence of the curvature dependent structure functions of the algebras of transformations between observers can be witnessed on the example of the Schwarzschild spacetime. Considering an observer located at $r_0 > r_S$ (in Schwarzschild coordinates, where $r_S$ is the Schwarzschild radius), think of two transformations:
\bse
\begin{align}
	Z\text{ parametrized by } t^\mu = \epsilon\delta^\mu_\tau,\quad l_\mu{}^\nu = 0,\\
	\bar Z\text{ parametrized by } \bar t^\mu = \epsilon\delta^\mu_r,\quad \bar l_\mu{}^\nu = 0,
\end{align}
\ese
which means one of them is an infinitesimal translation in time, while the other one is an infinitesimal translation in the radial direction. The algebra \eqref{algebraspacetime} tells us that the two observers which result from applying the above two transformations in different orders are located in the same position (since $\bar{\bar t}^\mu = 0$), but are boosted with respect to each other by a factor
\be
	\bar{\bar l}_{\mu\nu} = \epsilon^2
	\begin{bmatrix}
		0 & -\frac{r_S}{r_0^3} & 0 & 0 \\
		\frac{r_S}{r_0^3} & 0 & 0 & 0 \\
		0 & 0 & 0 & 0 \\
		0 & 0 & 0 & 0
	\end{bmatrix},
\ee
that is, a boost in the radial direction.

%~~~~~~~~~~~~~~~~~~~~~~~~~~~~~~~~~~~~~
\appendix
%~~~~~~~~~~~~~~~~~~~~~~~~~~~~~~~~~~~~~

%~~~~~~~~~~~~~~~
\section{Some properties of the metric in the adapted coordinates} \label{app:prop_norm_coor}
%~~~~~~~~~~~~~~~

When described in the Cartesian adapted coordinates, conditions \eqref{radialgauge} read
\be
	 (q_{IJ}(x)-\delta_{IJ})x^J = 0.
\ee
Acting with $\partial_K$, $\partial_L$ and $\partial_M$ in sequence we get
\be
	q_{IK,LM}(x) + q_{IL,KM}(x) + q_{IM,KL}(x) + q_{IJ,KLM}(x) x^J = 0,
\ee
so at the location of the observer we get
\be
	q_{IK,LM}(\sigma_0) + q_{IL,KM}(\sigma_0) + q_{IM,KL}(\sigma_0) = 0,
\ee
which means that the property $q_{I(J,MN)}(\sigma_0) = 0$, we used above indeed holds.

Using the above property and the fact that $q_{IJ,MN}$ is symmetric in both the first and last pair of its indices we can see that
\begin{align}
	q_{IJ,MN}(\sigma_0) &= - q_{MI,NJ}(\sigma_0) - q_{NI,MJ}(\sigma_0) \nonumber\\
	&= 2 q_{MN,IJ}(\sigma_0) + q_{JM,NI}(\sigma_0) + q_{JN,MI}(\sigma_0) \nonumber\\
	&= 2q_{MN,IJ}(\sigma_0) - q_{IJ,MN}(\sigma_0),
\end{align}
from which it follows that indeed $q_{IJ,MN}(\sigma_0) = q_{MN,IJ}(\sigma_0)$. Using this identity we obtain
\be\label{eq:app_riemann_norm_coor}
	R^K{}_{LIJ}(\sigma_0) = \Gamma^K{}_{LI,J}(\sigma_0) - \Gamma^K{}_{LJ,I}(\sigma_0) 
	= q^{KM}(\sigma_0)\left( q_{MI,LJ}(\sigma_0) - q_{MJ,LI}(\sigma_0) \right).
\ee
%~~~~~~~~~~~~~~~
\section{Properties of the vector field $Z$}\label{appendixdrugapochodnaZ}
%~~~~~~~~~~~~~~~

In this appendix we find a set of differential equations which are fulfilled by the vector fields $Z$.

The one-parameter group of diffeomorphisms generated by the field $g\mapsto Z[g]$
\be
	s \mapsto \psi_s^Z[g]
\ee
satisfies the equation
\be
	\frac{\text{d}}{\text{d}s} \psi_s^Z[g] = Z[ g_s ] \circ \psi_s^Z[g] 
\ee
where
\be
	g_s := \psi_s^Z[g]^* g 
\ee
As stated in \eqref{g0} and \eqref{radialgauge4d} we have
\bse\label{eq_g_s_0}
\begin{align}
	g_{s\,\mu\nu}\big|_\gamma &= \eta_{\mu\nu},\\
	\partial_\sigma g_{s\,\mu\nu}\big|_\gamma &= 0,\\
\intertext{where the notation $\big|_\gamma$ denotes the fact that the equations are satisfied along the worldline of the observer $\gamma$ and}
	g_{s\,r\mu} &= \eta_{r\mu}
\end{align}	
\ese
for all values of $s$. Therefore,
\bse\label{eq_derivative_g_s_0}
\begin{align}
	\left.\frac{\text{d}}{\text{d}s}\right|_{s=0} g_{s\,\mu\nu}\big|_\gamma &= 0,\\
	\left.\frac{\text{d}}{\text{d}s}\right|_{s=0} \partial_\sigma g_{s\,\mu\nu}\big|_\gamma = \partial_\sigma \left.\frac{\text{d}}{\text{d}s}\right|_{s=0} g_{s\,\mu\nu}\big|_\gamma &= 0, \\
	\left.\frac{\text{d}}{\text{d}s}\right|_{s=0} g_{s\,r\mu} &= 0.
\end{align}	
\ese
Using the fact that
\be
	\left.\frac{\text{d}}{\text{d}s}\right|_{s=0} g_{s\,\mu\nu} = {\cal L}_{Z[g]} g_{\mu\nu} 
	= Z^\rho \partial_\rho g_{\mu\nu}
	+ g_{\mu\rho} \partial_\nu Z^\rho
	+ g_{\rho\nu} \partial_\mu Z^\rho
\ee
we can rewrite \eqref{eq_derivative_g_s_0} in the form
\bse
\begin{align}
	(Z^\rho \partial_\rho g_{\mu\nu} + g_{\mu\rho} \partial_\nu Z^\rho + g_{\rho\nu} \partial_\mu Z^\rho)\big|_\gamma &= 0,
	\\
	\partial_\sigma(Z^\rho \partial_\rho g_{\mu\nu} + g_{\mu\rho} \partial_\nu Z^\rho + g_{\rho\nu} \partial_\mu Z^\rho)\big|_\gamma &= 0,
	\\
	{\cal L}_{Z} g_{r\mu} &= 0. 
\end{align}	
\ese
Taking into account \eqref{eq_g_s_0} we can simplify the first two equations
\bse\label{eq_derivative_g_0}
\begin{align}
	\eta_{\mu\rho} \partial_\nu Z^\rho\big|_\gamma + \eta_{\rho\nu} \partial_\mu Z^\rho\big|_\gamma = 0,
	\label{eq:first_derivative}
	\\
	Z^\rho \partial_\sigma\partial_\rho g_{\mu\nu}\big|_\gamma 
	+ \eta_{\mu\rho} \partial_\sigma\partial_\nu Z^\rho\big|_\gamma 
	+ \eta_{\rho\nu} \partial_\sigma\partial_\mu Z^\rho\big|_\gamma = 0.
\end{align}	
\ese
The latter of the above equations might be written equivalently as
\bse
\begin{align}
	-Z^\rho \partial_\sigma\partial_\rho g_{\mu\nu}\big|_\gamma 
	- \eta_{\mu\rho} \partial_\sigma\partial_\nu Z^\rho\big|_\gamma 
	- \eta_{\rho\nu} \partial_\sigma\partial_\mu Z^\rho\big|_\gamma &= 0,
	\\
	Z^\rho \partial_\nu\partial_\rho g_{\mu\sigma}\big|_\gamma 
	+ \eta_{\mu\rho} \partial_\nu\partial_\sigma Z^\rho\big|_\gamma 
	+ \eta_{\rho\sigma} \partial_\nu\partial_\mu Z^\rho\big|_\gamma &= 0,
	\\
	Z^\rho \partial_\mu\partial_\rho g_{\sigma\nu}\big|_\gamma 
	+ \eta_{\sigma\rho} \partial_\mu\partial_\nu Z^\rho\big|_\gamma 
	+ \eta_{\rho\nu} \partial_\mu\partial_\sigma Z^\rho\big|_\gamma &= 0.
\end{align}	
\ese
Adding the above equations we obtain
\be
	\eta_{\rho\sigma} \partial_\nu\partial_\mu Z^\rho\big|_\gamma =
	- \frac{1}{2} Z^\rho\partial_\rho \left( 
	\partial_\nu g_{\mu\sigma}+\partial_\mu g_{\sigma\nu}-\partial_\sigma g_{\mu\nu} 
	\right)\big|_\gamma =  - Z^\rho \partial_\rho\Gamma_{\sigma\mu\nu}\big|_\gamma
\ee 
or
\be
	\partial_\nu\partial_\mu Z^\rho\big|_\gamma =  - Z^\sigma \partial_\sigma\Gamma^\rho_{\phantom{\rho}\mu\nu}\big|_\gamma,
\ee
which means in particular that
\be\label{eq_second_derivative}
	\partial_\nu\partial_\mu Z^\rho(p) =  - Z^\sigma \partial_\sigma\Gamma^\rho_{\phantom{\rho}\mu\nu}(p),
\ee

%~~~~~~~~~~~~~~~~~
\section{Relation to the previous work on the subject of the spacetime radial gauge}\label{ZinDiracmatrix}
%~~~~~~~~~~~~~~~~~

In this appendix we show that the vector fields $Z$ discussed in section \ref{spacetimecase} are closely related to the considerations of the spacetime radial gauge in \cite{Bodendorfer:2015kua}. Also, we argue for the uniqueness of the solutions of equations \eqref{eq:determineZ}.

Let us first show that equation \eqref{eqNabla4d} implies
\be\label{eqRiemann4d}
	\nabla_r \nabla_r Z^\mu = R^\mu{}_{rr\nu}Z^\nu.
\ee
To this end, we write \eqref{eqNabla4d} in the form
\be\label{eq:app_dirac_g_nabla_Z}
	g(\nabla_r Z,\partial_\mu) = -g(\nabla_\mu Z,\partial_r),\quad \mu=\tau,r,A .
\ee
Next, we act with one more derivative operator $\nabla_r$ on both sides and find
\be
	g(\nabla_r\nabla_r Z,\partial_\mu) + g(\nabla_r Z,\nabla_r\partial_\mu) 
	= -g(\nabla_r\nabla_\mu Z,\partial_r) - g(\nabla_\mu Z,\nabla_r\partial_r).
\ee
The second term on the right-hand side drops because the coordinate lines of the radial coordinate are geodesic by definition meaning $\nabla_r\partial_r=0$. We move the second term on the left-hand side to the right and get
\be
	g(\nabla_r\nabla_r Z,\partial_\mu) 
	= - g([\nabla_r,\nabla_\mu] Z,\partial_r) - g(\nabla_\mu\nabla_r  Z,\partial_r) - g(\nabla_r Z,\nabla_r\partial_\mu),
\ee
where we have added and subtracted a term to introduce the commutator. The first term on the right-hand side can be reexpressed with the use of the Riemann tensor, while the second term can be rewritten using
\be
	- g(\nabla_\mu\nabla_r  Z,\partial_r) 
	= -\nabla_\mu\big( g(\nabla_r  Z,\partial_r)\big) + g(\nabla_r  Z,\nabla_\mu\partial_r).
\ee
The first term on the right-hand side vanishes since
\be
	g(\nabla_r  Z,\partial_r) = \nabla_r Z_r =  Z_{(r;r)}
\ee
and $Z_{(r;r)} = 0$ as a consequence of the equation \eqref{eq:app_dirac_g_nabla_Z}. Therefore, we get
\be
	g(\nabla_r\nabla_r Z,\partial_\mu) 
	= - R_{r\nu r\mu} Z^\nu + g(\nabla_r  Z,\nabla_\mu\partial_r - \nabla_r\partial_\mu) 
	= - R_{r\nu r\mu} Z^\nu,
\ee
where the last equality follows from the vanishing of the torsion. This concludes the proof of \ref{eqRiemann4d}.

Next, we introduce a decomposition of the vector field $Z$ into the parts normal and tangent to the surfaces of constant time   
\be
	Z = Z^\perp n + Z^a\partial_a,  
\ee
where $n$ is a normalized vector normal to the surface of constant time ($Z^\perp$ is `the lapse' and $Z^a$ is `the shift vector' of $Z$). The identity \eqref{eqRiemann4d} expressed in terms of $Z^\perp$ and $Z^a$ acquires the following form
\bse
\begin{align}
	\partial^2_r Z^\perp + 2 K_{ra}\partial_rZ^a - {}^{(3)}\!R_{rr} Z^\perp + 2K_{ra}K^{ra}Z^\perp &= 0,\\
	\partial_r(q_{ab}\partial_rZ^b + 2K_{ra}Z^\perp) &= 0,
\end{align}
\ese
where ${}^{(3)}\!R_{ab}$ is the Ricci tensor of the spatial metric $q_{ab}$ defining the intrinsic geometry of the constant time slices, while $K_{ab}$ is the extrinsic curvature of those slices. The above equations are equivalent to
\bse\label{eq:noncovZ}
\begin{align}
	\partial^2_r Z^\perp + 2 K_{ra}\partial_rZ^a - {}^{(3)}\!R_{rr} Z^\perp + 2K_{ra}K^{ra}Z^\perp &= 0,\\
	q_{ab}\partial_rZ^b + \partial_a Z^r + 2K_{ra}Z^\perp &= 0,
\end{align}
\ese
which shows that the fields $(Z^\perp,Z^a)$ lie in the kernel of the transposition of the $A$ block of the Dirac matrix introduced in Appendix A of \cite{Bodendorfer:2015kua}. 

The equations \eqref{eq:noncovZ} determine $(Z^\perp,Z)$ on each surface of constant time $\tau$ if we specify the following values: $Z^\perp(\tau,0,0,0)$, $Z^I(\tau,0,0,0)$ and $\partial_J Z^\perp(\tau,0,0,0)$, $\partial_J Z^I(\tau,0,0,0)$. Thus, to prove that the set of equations \eqref{eq:determineZfirst}--\eqref{eq:determineZthird} with initial conditions \eqref{initialconditionsforZ} have a single global solution, it is enough to show that $Z^\perp$, $Z^I$ and $\partial_J Z^\perp$, $\partial_J Z^I$ are uniquely determined along the world line of the observer. This is indeed the case -- first we determine $Z^\perp$ and $Z^I$ along the observer's world line with the use of \eqref{eq:determineZfirst} and the initial conditions, then we find $\partial_J Z^\perp$, $\partial_J Z^I$ using \eqref{eq:determineZsecond}.

%~~~~~~~~~~~~~~~~~~~~~~~~
\section*{Acknowledgements}
%~~~~~~~~~~~~~~~~~~~~~~~~

This work was partially supported by the Polish National Science Centre grant No.~2011/02/A/ST2/00300.

%~~~~~~~~~~~~~~~~~~~~~~~~
\providecommand{\href}[2]{#2}\begingroup\raggedright\endgroup

\end{document}